\documentclass[twoside,english,nofootinbib,preprintnumbers,aps,onecolumn,notitlepage]{revtex4-2}
\usepackage{euscript}
\usepackage{amsthm}
\usepackage{graphicx}
\usepackage{amsfonts}
\usepackage{amsmath}
\usepackage{amssymb}
\usepackage{mathtools}
\usepackage{fancyhdr}
\usepackage{epsfig}
\usepackage{color}
\usepackage{esint}
\usepackage{soul}
\usepackage{calc}
\usepackage{enumerate}
\usepackage{enumitem}
\usepackage{afterpage}
\usepackage{physics}
\usepackage[papersize={8.5in,11in}]{geometry}

\allowdisplaybreaks

\newcommand{\be}{\begin{equation}}
\newcommand{\ee}{\end{equation}}

\begin{document}
\title{Isolated horizons of the Hopf bundle structure transversal to the null direction, the horizon equations and embeddability in NUT-like spacetimes.}
\author{Denis Dobkowski-Ry{\l}ko}
	\email{Denis.Dobkowski-Rylko@ug.edu.pl}
 \affiliation{Institute of Theoretical Physics and Astrophysics, Faculty of Mathematics, Physics and Informatics, University of Gdansk, Wita Stwosza 57, 80-308, Gdańsk, Poland}
 \affiliation{Faculty of Physics, University of Warsaw, ul. Pasteura 5, 02-093 Warsaw, Poland}

\author{Jerzy Lewandowski}
	\email{Jerzy.Lewandowski@fuw.edu.pl}
	\affiliation{Faculty of Physics, University of Warsaw, ul. Pasteura 5, 02-093 Warsaw, Poland}
\author{Maciej Ossowski}
	\email{Maciej.Ossowski@fuw.edu.pl}
	\affiliation{Faculty of Physics, University of Warsaw, ul. Pasteura 5, 02-093 Warsaw, Poland}
\begin{abstract}    
Isolated horizons that admit the Hopf bundle structure $H\rightarrow S^2$ are investigated however, the null direction is allowed to be not tangent to the bundle fibers. 
The horizons are assumed to be axisymmetric. 
The geometry of such horizons is characterized  by an axially symmetric data set on a topological two-dimensional sphere: a metric tensor with a conical singularity at its poles, and a regular rotation scalar, in a nonextremal case. 
In the extremal horizon case one more regular function is needed. 
The horizon equations induced by Einstein's equations are imposed.   
The existence of regular extremal horizons satisfying the vacuum (with cosmological constant) equation of extremality, obtained from singular solutions on the sphere is pointed out.  
All horizons of the assumed topology and symmetry satisfying the $\Lambda-$vacuum Type D equation are derived. 
They are compared to the Killing horizons contained in the accelerated Kerr-NUT-(anti-) de Sitter spacetimes. 
If the cosmological constant takes special values determined by the other parameters the bundle fibers become tangent to the null direction.    
As an additional but also important result,  spacetimes of the topology $H\times \mathbb{R}$  locally isometric to the accelerated Kerr-NUT-(anti-)de Sitter spacetimes are constructed  for every value of the mass, Kerr, NUT parameters, the cosmological constant and the acceleration. 
When the acceleration parameter is not zero, the conical singularity can be removed whenever the NUT parameter does not vanish either.           

\end{abstract}

\date{\today}


\maketitle

\tableofcontents
\newpage
\section{Introduction}

\vspace{5mm}

The theory of isolated horizons provides a powerful framework for local study of black hole horizons and their generalizations.
Einstein equations, upon additional assumptions, induce constraints on the horizon geometry.
They allow one to formulate and prove - purely in terms intrinsic to the horizon - the properties analogous to those known from the globally defined, asymptotically flat spacetime theory of black holes, such as rigidity \cite{PhysRevD.97.124067, Dunajski2023}, no-hair \cite{GeometricCharacterizationsoftheKerrIsolatedHorizon, JerzyLewandowski_2003, typeD, Lewandowski_2003, Buk}, topologically spherical cross sections \cite{DOBKOWSKIRYLKO2018415,Dobkowski-Rylko:2018nti} . 
This is the case either when the horizon is assumed to be extremal (i.e. with vanishing surface gravity)  \cite{geometryhorizonsAshtekar_2002, higherDim, Lewandowski_2003, lucietti2009, Kunduri_2007, Lucietti2013} or if it is nonextremal, but the four-dimensional spacetime Weyl tensor is assumed to be of the Petrov Type D at the horizon \cite{GeometricCharacterizationsoftheKerrIsolatedHorizon, localnohairPhysRevD.98.024008}.
The isolated horizon's geometry, the equations it is supposed to satisfy and the horizon's topology can be introduced in a manner detached from the spacetime.
This is in contrast to the notion of embedded horizon.
Having obtained a family of isolated horizons with particular, desirable properties a natural question about embeddability arises.
This problem has led to new solutions to Einstein equations \cite{LO3PhysRevD.104.024022}, in turn containing - as will be demonstrated in the subsequent sections - new examples of isolated horizons.
We note that this approach is not the only possibility.
Locally, horizons have also been studied using Killing spinors, leading to alternative classification of admissible geometries on sections \cite{Cole_2018}.

In the current work, we introduce and explore horizons of a new type, specifically with a new topology.
We consider three-dimensional isolated horizons that are compatible with four-dimensional spacetimes.
The simplest topology of an isolated horizon $H$ is the product $H=S \times \mathbb{R}$, where $S$ is a compact 2-manifold and fiber $\{s\}\times \mathbb{R}$ over any $s \in S$ is null with respect to the degenerate geometry of the horizon. 
Such horizons are contained in Kerr and Kerr-de Sitter spacetimes, with $S$ being a topological $2$-sphere. 
The product horizons with $S$ of higher genus are to be found in toroidal or hyperbolic Schwarzschild-(anti-)de Sitter spacetimes. 
An example of nonproduct topology is a horizon of a $U(1)$ principal fiber bundle structure $H \rightarrow S$, where the fibers coincide with the null curves defined by the degenerate metric tensor of $H$. 
Such horizons are contained in Kerr-NUT-(anti-)de Sitter or Taub-NUT-(anti-)de Sitter spacetimes.
They were also considered and studied locally, without embedding \cite{hopf}. 

The horizons we consider here also have the topology of the Hopf bundle 
\begin{equation}
    H=S^3 \rightarrow S^2
\end{equation}
($S^3$ and $S^2$ are topological spheres) however, the fibers are not null with respect to the degenerate metric tensor on $H$.
In other words, the null curves in $H$ are transversal to the bundle fibers.
We present a construction of a general isolated horizon geometry of this topology that is invariant with respect to rotations and the action of the structure group. 
The motivation for the introduction of such horizons comes from the study of recently constructed globally defined Kerr-NUT-(anti-)de Sitter spacetimes of the topology $H\times \mathbb{R}$ free of conical singularities \cite{LO3PhysRevD.104.024022}.

{
\color{black}
While the solutions to the Petrov Type D equation with conical singularity can be used to construct singularity-free horizons with Hopf bundle structure, this is not necessary.
Without performing the topological identification one could instead construct a horizon with a product topology $S \times\mathbb{R}$, where $S$ is only diffeomorphic to the sphere due to the conical singularities. 
Such horizons could be embedded as a Cauchy horizon in a spacetime possessing a cosmic string and have been of interest for a long time in the context of a cosmic string snapping and a subsequent generation of gravitational waves \cite{SAHayward_1989,JPodolský_2000}.
Recently there has also been a resurgence of mathematical interest in spacetimes with conical singularities in the field of the black hole thermodynamics.
It has been shown that a consistent formulation of the thermodynamics of spacetimes with conical singularity is possible, with contributions from the cosmic string \cite{Appels2017,Bordo2020}.
}

This paper is organized as follows.
In Sec \ref{horizongluing} we define the isolated horizons with a structure of a principle bundle with the action generated by a Killing vector field transversal to the null direction. 
We point out the connection to extreme horizons.
In Sec \ref{sec:petrovD} we introduce the Petrov Type D equation without the regularity condition ensuring the lack of the conical singularity.
In Sec \ref{sec:sols} we solve it and derive a new class of solutions corresponding to horizons with conical singularity.
Finally, in Sec \ref{sec:embeddability} we address the problem of the existence of spacetime in a class of accelerated Kerr-NUT-(anti-)de Sitter solutions in which our horizons are embeddable.
Furthermore, we illustrate that the horizons in nonsingular accelerated Kerr-NUT-(anti-)de Sitter spacetimes (i.e. after performing the Misner gluing along suitable Killing vector field) admit precisely the structure of isolated horizons introduced in Sec {\ref{horizongluing}}.

\section{Isolated horizons of the Hopf bundle structure admitting conical singularities}
\label{horizongluing}

An (unembedded) isolated horizon $H$ is a three-dimensional surface equipped with the following:
\begin{enumerate}[label=(\roman*)]
    \item a degenerate metric tensor $g$ of a signature $0,+,+$;
    \item a torsion-free covariant derivative $\nabla$ on $T(H)$ which satisfies pseudo metricity condition:
\begin{align}\label{metricity}
    \nabla_a g_{bc}=0;
\end{align}
\item a nowhere vanishing vector field $\ell$ (defined up to a constant scaling):
\begin{align}\label{nullell}
    \ell^a g_{ab}=0,
\end{align}
such that the following equality is satisfied:
\begin{align}
    [\mathcal{L}_\ell,\nabla_a]=0.
\end{align}
\end{enumerate}
From (\ref{metricity}) and (\ref{nullell}) follows that the degenerate metric $g$ is Lie-dragged along $\ell$, namely
\begin{align}
\mathcal{L}_\ell g_{ab} = 0.
\end{align}
On a horizon we define the rotation 1-form potential $\omega$ via:
\begin{align}
\label{eq:omega-def}
    \nabla_a \ell^b = \omega_a \ell^b,
\end{align}
whereas its contraction with $\ell$ determines surface gravity $\kappa^{(\ell)}$:\begin{align}
\label{eq:kappa-def}
    \ell^a\omega_a=:\kappa^{(\ell)}.
\end{align}
Notice, that surface gravity depends on the scaling of $\ell$, which results in
\begin{align}
\kappa^{(c_0\ell)}=c_0\kappa^{(\ell)}, \ \ \ \ \ \ \  \text{ for } \ c_0=\text{const}.
\end{align}
{
\color{black}
From Einstein equations for the $\Lambda-$vacuum  (alternatively one may assume the dominant energy condition for the energy-stress tensor with the cosmological constant included) it follows that surface gravity is constant, whereas its vanishing or nonvanishing means that $H$ is an extremal or a nonextremal isolated horizon, respectively.}
Furthermore, the degenerate metric $g$ uniquely determines the degenerate volume 2-form $\eta_{AB}$, where $A$ and $B$ are indices of tensors defined on $S^2$, such that its pullback to every spacelike two-dimensional slice of $H$ is the area 2-form. 
It is then used to introduce rotation pseudoscalar $\Omega$, such that
\begin{align}\label{OmegaDef}
    d\omega_{AB} =: \Omega \eta_{AB}.
\end{align}

In the literature, one usually  encounters isolated horizons of a cylinder topology $H = S\times \mathbb{R}$,
where $S$ is a two-dimensional connected space of null curves in $H$.
Recently, isolated horizons were also considered on a nontrivial  $U(1)$ bundle \cite{hopf}, where the flow of the null vector field $\ell$ was assumed to coincide with the action of $U(1)$ on $H$. 
In the current work we further generalize the topology of the isolated horizon by considering a horizon $H$ of the Hopf bundle structure:
\begin{align}
 \Pi: H \rightarrow S^2, 
\end{align}
where,  the null vector field $\ell$ is not assumed to be tangent to the fibers.
Below, we provide a detailed construction of a generic isolated horizon using the elements of intrinsic geometry of $H$, that is a degenerate metric tensor $g$ and rotation 1-form potential $\omega$.
Consequently, in a nonextremal case, vacuum Einstein equations determine  the connection $\nabla$. 

As a starting point, consider two isolated horizons:
\begin{align}\label{twoHorizons}
\mathcal{H}=S^2\backslash \{p_{\pi}\}\times S^1 && \text{and} &&
\mathcal{H}'=S^2\backslash \{p_{0}\}\times S^1.
\end{align}
On $\mathcal{H}$ we consider coordinates $(\theta, \varphi, v)$ whereas on $\mathcal{H}'$ we have $(\theta', \varphi', v')$, which provide the following metrics on the two horizons:
\begin{align}
\label{Hmetric}
    g_{ab} dx^a dx^b =U^2(\theta)\bigg[ d\theta^2 +f^2(\theta) \sin^2{\theta} \Big(\frac{1}{f(0)}d\varphi + \Big( \frac{1}{f(\pi)}-\frac{1}{f(0)}\Big)dv \Big)^2\bigg]
\end{align}
and
\begin{align}\label{H'metric}
    g'_{ab} dx'^a dx'^b = U^2(\theta')\bigg[d\theta'^2 +f^2(\theta') \sin^2{\theta'} \Big(\frac{1}{f(\pi)}d\varphi' + \Big( \frac{1}{f(0)}-\frac{1}{f(\pi)}\Big)dv' \Big)^2\bigg], 
\end{align}
where $\theta \in [0,\pi)$, $\theta' \in (0,\pi]$, $\phi,v,\phi',v'\in[0,2\pi)$ and the conformal factor $U^2$ is smooth everywhere on the horizon. 
Notice that both metrics $g$ and $g'$ admit conical singularities at $\theta=\pi$ and $\theta'=0$, respectively, unless
$f(0) = f(\pi)$. 
Here we consider a general case when such an equality does not necessarily occur. 
Next, we introduce a gluing of the two charts defined by the following transformation law:
\begin{align}\label{transformation}
   \theta=\theta', && \varphi=-\varphi', && v=v'-\phi', &&\text{for} \ \ \ \theta,\theta'\neq 0,\pi.
\end{align}
Therefore, the two charts $(\theta, \varphi, v)$ and $(\theta', \phi', v')$ together with the transition map (\ref{transformation}) provide an atlas on $H$. Metrics (\ref{Hmetric}) and (\ref{H'metric}) uniquely define a smooth, symmetric, rank 2 tensor on a three-dimensional null surface $H$. 
The transition map (\ref{transformation}) generates a Hopf bundle structure, where the $U(1)$ action is determined by the vector field $\partial_v$ and $\partial_{v'}$, respectively. 
Notice, however, that it does not coincide with the null vector fields $\ell$. 
We may choose $\ell$ to be of the form:
\begin{align}
\label{eq:ell-transversal-construction}
\ell&=f(\pi)\partial_v+\Big(f(\pi)-f(0)\Big)\partial_\varphi\nonumber\\
&=f(0)\partial_{v'}+\Big(f(0)-f(\pi)\Big)\partial_{\varphi'},
\end{align}
which is equivalent to fixing its normalization.
Next, on $H$ we define a rotation 1-form potential $\omega$ such that:
\begin{align}
    \ell  \ \lrcorner \ \omega = \kappa = \text{const.} && \text{and} && \mathcal{L}_\ell \omega = 0.
\end{align}
It follows that in the coordinates $(\theta,\phi,v)$ the rotation 1-form potential is of the form:
\begin{align}
\label{omegaFinalpi}
    \omega = \frac{2\kappa}{f(\pi)+f(0)}\bigg(dv-\frac{1}{2}d\varphi\bigg)+h(\theta)\bigg(\frac{1}{f(0)}d\varphi + \Big( \frac{1}{f(\pi)}-\frac{1}{f(0)}\Big)dv \bigg)+dw(\theta),
\end{align}
where $h(\theta)$ is {\it{a priori}} an arbitrary function.
Requiring $\omega$ to be well-defined at $\theta=0$ results in the following condition:
\begin{align}\label{omega_0}
    \omega_\phi (0) = \frac{h(0)}{f(0)} - \frac{\kappa}{f(\pi)+f(0)}=0.
\end{align}
Similarly, for the primed coordinates $(v',\theta', \varphi')$ we have:
\begin{align}
    \omega = \frac{2\kappa}{f(\pi)+f(0)}\bigg(dv'-\frac{1}{2}d\varphi'\bigg)-h(\theta)\bigg(\frac{1}{f(\pi)}d\varphi' + \Big( \frac{1}{f(0)}-\frac{1}{f(\pi)}\Big)dv' \bigg)+dw(\theta')
\end{align}
and, to be well-defined at $\theta$, it needs to satisfy:
\begin{align}\label{omega_pi}
\omega_{\varphi'}(\pi)=-\frac{h(\pi)}{f(\pi)}-\frac{\kappa}{f(\pi)+f(0)}=0.
\end{align}
Instead of the rotation 1-form potential $\omega$, one may operate with the rotation pseudoscalar 
$\Omega$, defined in (\ref{OmegaDef}), where:
\begin{align}
    \eta &= U^2(\theta)f(\theta)\sin\theta d\theta\wedge\bigg( \frac{1}{f(0)}d\varphi + \Big( \frac{1}{f(\pi)}-\frac{1}{f(0)}\Big)dv\bigg)\nonumber\\
    &= -U^2(\theta')f(\theta')\sin\theta'd\theta'\wedge\bigg( \frac{1}{f(\pi)}d\varphi' + \Big( \frac{1}{f(0)}-\frac{1}{f(\pi)}\Big)dv'\bigg).
\end{align}
Furthermore, the rotation pseudo 
scalar $\Omega$ determines the function $h(\theta)$ up to an additive constant $c$:
\begin{align}
    h(\theta)= \int\Omega(\theta) U^2(\theta)f(\theta)\sin\theta d\theta=:h_0(\theta)+c.
\end{align}
We find the integration constant $c$ from the first constraint (\ref{omega_0}):
\begin{align}
    c=\frac{f(0)\kappa}{f(\pi)+f(0)}-h_0(0).
\end{align}
Then, the second condition (\ref{omega_pi}) becomes a constraint on surface gravity $\kappa$, which is of the form
\begin{align}
\label{ConditionKappa}
\kappa = h_0(0)-h_0(\pi).
\end{align}
Therefore, it is the surface gravity $\kappa$ which defines an appropriate way of gluing two horizons, $\mathcal{H}$ and $\mathcal{H}'$. 
Consequently, the horizon $H$ defined by the two charts, $(\theta,\varphi,v)$ and $(\theta',\varphi',v')$, is diffeomorphic to $S^3$.

The construction (\ref{twoHorizons}-\ref{ConditionKappa}) described above, 
may be performed for every 2-sphere metric
\begin{align}
\label{genMetric}
    g_{AB}dx^A dx^B = g_{xx} dx^2 + g_{\psi \psi}d\psi^2
\end{align}
with {\it{a priori}} two conical singularities at the poles, $x=\pm 1$, 
and arbitrary function
\begin{align}
 \Omega=\Omega(x)   
\end{align} that is regular on the entire sphere (including the poles). Metric singularities may be removed at each pole separately (but not simultaneously) by introducing the following transformations:
\begin{align}
\label{psi-phi}
    \varphi=\frac{1}{\beta}\psi, &&
    \varphi' =\frac{1}{\beta'}\psi,
\end{align}
and setting appropriate values for $\beta$ and $\beta'$ in a similar fashion as described above, where $\beta$ and $\beta'$ correspond to $1/f(0)$ and $1/f(\pi)$ so that 
\begin{equation}
\label{eq:phi-in-0-2pi}
    \varphi,\varphi' \in [0,2\pi).
\end{equation}
If the resulting $\kappa\neq 0$, then the term $\dd w(x)$ in $\omega$ can always be set to be zero, due to transformations $v=v' + \tilde{w}(x)$. 
In the case $\kappa=0$, we additionally need
\begin{align}\label{wtheta}
    w=w(x)
\end{align} 
to be regular on the entire sphere.
\bigskip 

Several general remarks about our construction are in order.

\noindent{\bf Remark 1: nongeneric sub-cases.}  Given the Hopf bundle $H\rightarrow S^2$:
\begin{itemize}
    \item every pair $(g_{AB},\omega_A)$ defined on the entire $S^2$ without conical singularities corresponds to an extremal horizon  structure  ($
    \kappa=0$) such that $\ell$ is tangent to the fibers of the Hopf bundle  $H\rightarrow S^2$. Notice, however, that such  structure in general does not satisfy the vacuum extremal equation introduced in Remark 3 below. 
    \item if $g_{AB}$ is regular on the entire $S^2$,  then $\ell$ is tangent to the fibers of the  bundle.
\end{itemize}
\bigskip

\noindent{\bf Remark 2: the cyclic symmetry.} 
Our construction is modeled on the Hopf bundle where  $S^3$ also admits a part of the $SU(2)$ group structure.
Namely, we assume that there exists right $U(1)$ action on $S^3$ generated by $\partial_v$ and that the corresponding left $U(1)$ action is generated by 
\begin{equation}
\label{rotationsym}
\partial_\phi + \frac{1}{2}\partial_v = -\bigg(\partial_{\phi'}+\frac{1}{2}\partial_{v'}\bigg).
\end{equation}
Furthermore, we assume that both of them are symmetries of the degenerate horizon metric and that the vector field $\ell$ spanning the null direction is their linear combination.

In the generic case the orbits of $\ell$ are dense in the orbits spanned by $\ell$ and $\partial_v$.

\noindent{\bf Remark 3: extremal vacuum horizons.}
Given a $2$-dimensional manifold endowed with a metric tensor $g_{AB}$ and a differential $1$-form $\omega_{AB}$, the vacuum  extremal horizon equation with a cosmological constant $\Lambda$ reads
\begin{equation}
D_{(A} \omega_{B)} + \omega_A\omega_B - \frac{1}{2}R_{AB} + \frac{1}{2}\Lambda g_{AB}=0, \label{ext}
\end{equation}
where $D_A$ is the torsion-free, metric connection corresponding to $g_{AB}$ and $R_{AB}$ is its Ricci tensor. 
Extremal isolated horizon $H$ is vacuum extremal if there is a constant $\Lambda$ such that the vacuum extremal horizon equation is satisfied for every local section transversal to the null vector $\ell$ by the pullback of the degenerate metric tensor and the rotation $1$-form potential $\omega$. 
A new conclusion coming from this section is that a regular vacuum  extremal horizon structure on the Hopf bundle $H$ is obtained by implementing our very construction to any solution of the vacuum extremal horizon equation defined on $S^2$ such that
\begin{enumerate}
    \item $(g_{AB},\omega_A)$ is axisymmetric,    
    \item $g_{AB}$ is regular except the poles where it can have conical singularities,
    \item $\omega_A$ is regular everywhere.
\end{enumerate}
Since all the axisymmetric solutions of (\ref{ext}) on $S^2$ with conical singularities are known  \cite{Lucietti2013} it is easy to check by inspection which of them satisfy our conditions - we anticipate that all of them do. 
\bigskip

In the following section, we will introduce the Type D equation that comes from Einstein equations and is imposed on scalar $K$ characterizing the degenerate metric of $H$ and rotation pseudoscalar $\Omega$. Although we do not explicitly bring up other elements of the horizon connection $\nabla$, one may, however, retrieve them from $(g_{AB}, \Omega)$ in the nonextremal case, or they are subject to other equations in the extremal case \cite{geometryhorizonsAshtekar_2002, Szereszewski2019,Li_2016}.

\section{Petrov Type D equation}
\label{sec:petrovD}
Given an isolated horizon $H$ and its geometry $g$ and $\nabla$, and the null symmetry generator $\ell$ the Petrov Type D vacuum horizon equation is imposed on the degenerate metric tensor $g$ and the rotation pseudoscalar $\Omega$. 
This equation is also satisfied by every isolated horizon that is contained as a Killing horizon (to the second order) in $\Lambda$-vacuum spacetime such that its Weyl tensor is of the Petrov Type D on $H$. 
It is also satisfied by every vacuum extremal horizon regardless of the Weyl tensor. 
The Petrov Type D equation was investigated assuming various topologies of the isolated horizons. 
In particular, all the axisymmetric solutions on the  horizons of the topology $\mathbb{R}\times\mathbb{S}_2$ were found in \cite{GeometricCharacterizationsoftheKerrIsolatedHorizon} for the vanishing cosmological constant $\Lambda$ and in \cite{typeD,localnohairPhysRevD.98.024008},  for $\Lambda\neq 0$, where their embeddability in the Kerr-(anti-)de Sitter spacetimes has been discussed. 
The conclusion was the uniqueness of the nonextremal Kerr-(anti-)de Sitter isolated horizons. 
For the case of two vacuum Type D isolated horizons of the topology $S\times \mathbb{R}$, which form a bifurcated horizon the axial symmetry was shown \cite{PhysRevD.97.124067}.
Next, the general solution was derived for the higher genus sections of the isolated horizons \cite{DOBKOWSKIRYLKO2018415} and found  to be nonrotating and of a constant scalar curvature $K$.
The next step was to consider the Type D equation on the horizons of the Hopf bundle structure \cite{hopf}. 
However, it was assumed that the null vector field $\ell$ of the horizon is tangent to the fibers of the bundle. 
Furthermore, rotational symmetry with respect to the vector field (\ref{rotationsym}) was assumed. 
Upon those assumptions, all of the solutions of the Type D equations were found. 
A subfamily of such isolated horizons is embeddable in the Taub-NUT-(anti-)de Sitter spacetimes or, assuming the following relation between the cosmological constant and the parameters of the horizon:
\begin{align}\label{LambdaStare}
    \Lambda = \frac{3}{a^2+2l^2+2r^2_H}
\end{align}
in the Kerr-NUT-(anti-)de Sitter spacetimes \cite{LO}. 
In Sec \ref{sec:Differentiability condition} we generalize the condition to the case of accelerated Kerr-NUT-(anti-)de Sitter spacetimes.

The horizons with structure of a Hopf bundle over Riemann surfaces with higher genus have also been studied \cite{LO4genus}. 
Similar to the trivial case the Type D equation implies that $K$ is constant, but now the horizons are allowed to rotate with constant $\Omega$.
All such horizons have been found to be embeddable into toroidal or hyperbolic Taub-NUT-(anti-)de Sitter spacetimes.

Given a $3$-dimensional  isolated horizon $H$ of the null vector field $\ell$ and  geometry $(g,\nabla)$, the Petrov Type D equation is defined on every two-dimensional {slice} of $H$ transversal to $\ell$. It   involves  the rotation pseudoscalar $\Omega$ defined by (\ref{OmegaDef}) and the Gaussian curvature $K$ of the 2-metric $g_{AB}$ induced on the slice
\begin{align}
R_{AB}=:K g_{AB},
\end{align}
where $R_{AB}$ is the Ricci tensor of $g_{AB}$. 
In the current work, we will  study and solve the Type D equation on a topological 2-sphere $S^2$, assuming the axial symmetry of $g_{AB}$ and $\Omega$, and allowing for conical singularities of $g_{AB}$.
Next, we  will compare the resulting family of solutions with those defined by horizons in the generalized Plebański-Demiański spherical black hole spacetimes known from the literature \cite{griffiths_podolsky_2009}.  

Since the Type D equation is expressed in terms of the complex structure defined by the metric $g_{AB}$, we start by introducing the complex null coframe $m_A$, that is:
\begin{align}
    g_{AB}=m_A \bar m_B + m_B \bar m_A, && \eta_{AB} = i(\bar m_A m_B - \bar m_B m_A).
\end{align}
It is convenient to work with the coordinates adapted to the symmetry, namely:
\begin{align}\label{2metric}
    g_{AB}dx^A dx^B = R^2\bigg(\frac{1}{P^2(x)}dx^2 +P^2(x) d\psi^2\bigg),
\end{align}
 where $x\in [-1;1]$ and we refer to $R$ as the area radius and $P^2$ - the frame coefficient \cite{localnohairPhysRevD.98.024008}. 
 {\color{black} 
 For these metrics, we will apply the topologically non-trivial horizon geometry construction method presented in the previous chapter. 
 Consequently, the range of dependence  of the $\psi$-coordinate is irrelevant at this point, and we consider $\psi$ taking values in an a priori undetermined interval. 
 Proper rescalings of that  coordinate will be adjusted in each of the poles independently  (see (\ref{genMetric}-\ref{eq:phi-in-0-2pi})).
 }The complex frame vector and its dual may be written in the following form:
 \begin{align}
     m^A\partial_A =\frac{1}{R\sqrt{2}}\bigg(P(x) \partial_x+i\frac{1}{P(x)}\partial_\psi\bigg), && \bar m_A dx^A = \frac{R}{\sqrt{2}}\bigg( \frac{1}{P(x)}dx-iP(x)d\psi\bigg).
 \end{align}
 In contrast to \cite{hopf} we  consider here a more general structure and relax the condition on the frame coefficient $P^2$, by requiring it to be vanishing only on the poles, that is
 \begin{align}
 \label{P^2poles}
     P^2(1) = 0 = P^2(-1).
 \end{align}
{
\color{black} That condition implies that a possible singularity in the poles is at most conical.
}
The condition for lack of conical singularity in terms of the frame coefficient $P^2$ reads
\begin{equation}
\label{lackconical}
    \partial_x P^2 \eval_{x=\pm 1} = \mp 2.  
\end{equation}
{\color{black} This condition, if satisfied,  implies the continuity of the metric tensor (\ref{2metric}) on the sphere.}
However, if the conical singularities are present at the poles, then the condition is relaxed to 
\begin{equation}
\label{con}
 \partial_x P^2\eval_{x=-1} >0,    \ \ \ \ \     \partial_x P^2\eval_{x=1} <0. 
\end{equation}  
Then, by changing the range of the coordinate $\psi$ or equivalently by its rescaling, one may get rid of the conical singularity at any one of the poles and retrieve  the corresponding condition in (\ref{lackconical}). 
{\color{black} Hence, we will assume (\ref{con}) throughout this section.}
Consequently, the 2-metric (\ref{2metric}) itself generically will not be differentiable nor continuous simultaneously on both of the poles. It will lead us to a new four parameter family of solutions to the Type D equation. 

Finally, the regularity condition that has to be satisfied by the function $\Omega$ on the poles reads
\begin{equation}
  g^{AB} D_A \Omega D_B \Omega = 0  
\end{equation}
and in our coordinates it is of the form
\begin{equation}\label{Omcon}
 \left( \left(\frac{d\Omega}{dx}\right)^2 P^2\right)\eval_{x=\pm 1} = 0.  
\end{equation}
Every solution $(g,\Omega)$ that satisfies the conditions (\ref{P^2poles}), (\ref{con}) and (\ref{Omcon}) gives rise to an isolated horizon structure on the Hopf bundle $H$, according to the construction presented in the previous section,
 {\color{black} that is at least continues at the points of the fibers corresponding to the poles of the sphere. However, the actual horizons determined below by those conditions and by the Petrov Type D equation   will be everywhere smooth and even analytic.}

The Petrov Type D equation expressed in the null-frame is of the form:
\begin{align}\label{typeDcov}
    \bar m^A \bar m^B D_A D_B \bigg( K-\tfrac{\Lambda}{3}+ i\Omega \bigg)^{-\frac{1}{3}} = 0,
\end{align}
where $D_A$ is the torsion-free, metric covariant derivative associated with $g_{AB}$ and the term in the bracket is nonvanishing, that is:
\begin{align}
    K-\tfrac{\Lambda}{3}+ i\Omega \neq 0.
\end{align}
In the introduced coordinates (\ref{2metric}), the Type D equation (\ref{typeDcov}) may be simply written as:
\begin{align}\label{typeD}
    \partial_x^2\Psi_2^{-\tfrac{1}{3}} =0,
\end{align}
 where $\Psi_2$ is of the form:
 \begin{align}
     \Psi_2 = -\tfrac{1}{2}(K+i\Omega) +\tfrac{\Lambda}{6}
 \end{align}
 and is the only nonzero invariant component of the Weyl tensor of the Type D. The solution to eq. (\ref{typeD}) may be written as:
 \begin{align}\label{psic1c2}
     \Psi_2=\frac{1}{(c_1 x+c_2)^3},
 \end{align}
 where $c_1$ and $c_2$ are complex constants. Expressing the Gaussian curvature $K$ in terms of the frame coefficient $P^2$ and its derivatives yields:
 \begin{align}
     \frac{1}{(c_1 x+c_2)^3} = \frac{1}{4R^2}\partial_x^2 P^2 -\frac{1}{2}i\Omega +\frac{\Lambda}{6}.
 \end{align}
 The above complex equation may be split into its real and imaginary parts:
 \begin{align}\label{typeDexplicit}
    \text{Re}\bigg[\frac{1}{(c_1 x+c_2)^3}\bigg] &= \frac{1}{4R^2}\partial_x^2 P^2 +\frac{\Lambda}{6},\\ \label{typeDim}
    \text{Im}\bigg[\frac{1}{(c_1 x+c_2)^3}\bigg] &=-\frac{1}{2}\Omega.
 \end{align}
 In the following chapter we find solutions to the above equations assuming that the metric coefficient $P^2$ vanishes at the poles (\ref{P^2poles}) and satisfies inequalities (\ref{con}).

\section{Solutions to the Type D equation  with conical singularities}

\label{sec:sols}

\subsection{Special solution for vanishing $c_1$}
 \label{sec:special-solution}
First, consider the case when the complex parameter $c_1$ vanishes. Consequently, Eq. (\ref{typeDexplicit}) becomes
\begin{align}
    A = \partial_x^2 P^2
\end{align}
for some real parameter $A$. Its solution is of the form
\begin{align}
    P^2(x) = \tfrac{1}{2}Ax^2+Bx+C,
\end{align}
where $B$ and $C$ are real constants. Requiring the continuity of the frame coefficient $P^2$ (\ref{P^2poles}) yields
\begin{align}
P^2(x) = C(1-x^2).
\end{align}
Using the above expression we obtain the following metric tensor $g_{AB}$:
\begin{align}
    g_{AB}dx^Adx^B &= R^2 \bigg(\frac{1}{P^2(x)}dx^2 +P^2(x) d\psi^2\bigg)\nonumber\\
    &= {R'^2} \bigg(\frac{1}{P'^2(x)}dx^2 +P'^2(x) d\varphi^2\bigg),
\end{align}
where we rescaled the area radius and angular coordinate, namely $R'^2=R^2/C$ and $d\varphi=C d\psi$. 
Therefore, the metric remains of the form (\ref{2metric}) with the frame coefficient
\begin{align}\label{Pspecial}
    P'^2(x) = 1-x^2
\end{align}
which on the poles not only satisfies the continuity (\ref{P^2poles}) but also the differentiability condition
\begin{align}
    \partial_x P'^2 \Big |_{x=\pm 1} = \mp 2.  
\end{align}
Consequently, the 2-metric $g_{AB}$ lacks conical singularities and is differentiable at the poles $x=\pm 1$. 
Furthermore, from the imaginary part of the Type D equation (\ref{typeDim}) and the vanishing of the complex constant $c_1$ follows that the rotation pseudoscalar $\Omega$ is constant:
\begin{align}
    \Omega=\text{const}.
\end{align}
{\color{black} Since {$g_{AB}$} is regular on the entire sphere, the solution falls in the nongeneric case of Remark 1, either the first built if $\Omega=0$ or the second built is $\Omega\not=0$. In the latter case the null flow is tangent to the fibers of the Hopf bundle and such solutions were considered in \cite{hopf}, however, the Killing vector has not been normalized properly there. We correct this error in Appendix.}

\subsection{Generic solution for nonvanishing $c_2$}
\label{sec:Generic solution for nonvanishing}
A general axisymmetric solution to the real part of the Petrov Type D equation (\ref{typeDexplicit}) is of the form:
\begin{align}
P^2 &= 2R^2 \text{Re}{\big[\tfrac{1}{c_1^2(c_1 x+c_2)}\big]} -\tfrac{1}{3}R^2\Lambda x^2 +a_0x+b_0,
\end{align}
where $a_0$ and $b_0$ are real integration constants. 
Introducing a new parametrization
\begin{align}
\label{par1}
A :&= \frac{c_1\bar c_2+\bar c_1 c_2}{c_1\bar c_1}, && D:= \frac{1}{R^2} \Big( A a_0 +b_0 \Big) - \frac{\Lambda}{3}B,\nonumber\\
B :&= \frac{c_2\bar c_2}{c_1 \bar c_1}, && E:= \frac{c_1^3+\bar c_1^3}{c_1^3\bar c_1^3} +\frac{1}{R^2}\Big( B a_0+Ab_0\Big),\\
C:&= \frac{a_0}{R^2}-\frac{\Lambda}{3}A, &&
F:= \frac{c_1^2 c_2+\bar c_1^2\bar c_2}{c_1^3\bar c_1^3} +\frac{b_0}{R^2} B,\nonumber
\end{align}
yields a more compact expression for the frame coefficient
\begin{align}
P^2&= R^2\frac{-\frac{\Lambda}{3}x^4+Cx^3+ Dx^2+Ex +F}{x^2+A x+B}.
\end{align}
We demand that it satisfies Eq. (\ref{P^2poles}) on the poles. Therefore, we obtain two constraints for parameters $C$, $D$, $E$ and $F$, that is
\begin{align}
    P^2(1) = -\frac{\Lambda}{3}+C+D+E+F&=0,\\
P^2(-1) = -\frac{\Lambda}{3}-C+D-E+F&=0,
\end{align}
from which it follows that
\begin{align}
F&=\frac{\Lambda}{3}-D && \text{and} && E=-C.
\end{align}
Consequently, we write the frame coefficient $P^2$ as an expression manifestly satisfying constraints (\ref{P^2poles}), namely:
\begin{align}
P^2&=R^2\frac{-\frac{\Lambda}{3}x^2+Cx+D-\frac{\Lambda}{3}}{x^2+A x+B}(x^2-1).
\end{align}
Next, we change parameter $D$ to $D'+\tfrac{\Lambda}{3}$ so that
\begin{align}
P^2&=R^2\frac{-\frac{\Lambda}{3}x^2+Cx+D'}{x^2+A x+B}(x^2-1),
\end{align}
and we drop the prime to make the expression more elegant:
\begin{align}\label{Pfinal}
P^2&=R^2\frac{-\frac{\Lambda}{3}x^2+Cx+D}{x^2+A x+B}(x^2-1).
\end{align}
Moreover, we solve the imaginary part of the Type D equation (\ref{typeDim}), to find the rotation pseudoscalar $\Omega$:
\begin{align}
\label{OmegaGen}
    \Omega &= \bigg(x^3 \Big (-3 AC + 3A^3C - 9ABC + 6 (1 + B)D + A^4 \Lambda - 4A^2B \Lambda + 2B (1 + B) \Lambda - A^2 (3D + \Lambda) \Big) \nonumber\\
    &+x^2 \Big(9A^2BC - 18 B (1 + B)C + 9AD - 9ABD + 3A^3B \Lambda - 3 AB (1 + 3 B) \Lambda\Big) \nonumber\\
    &+x \Big(-9 ABC + 9 AB^2C + 9A^2D - 18B(1 + B)D +3A^2B^2 \Lambda - 6B^2 (1 +B) \Lambda\Big)\nonumber\\
   & -3A^2BC + 6B^2 (1 +B)C +3A^3D - 9ABD-3AB^2D +A(-1+B)B^2 \Lambda\bigg) \nonumber\\
   &\bigg/\Big(3\sqrt{4B-A^2}\big(B+x (A+x)\big)^3\Big).
\end{align}
Notice that for the frame coefficient $P^2$ (\ref{Pfinal}) and rotation pseudoscalar $\Omega$ (\ref{OmegaGen}) to be well-defined the parameters $A$, $B$, $C$ and $D$ must satisfy the following conditions:
\begin{align}
&\text{for } \Lambda<0: \ \  A^2-4B<0 \ \ \ \wedge \ \ \ \tfrac{3}{2\Lambda}(C\pm\sqrt{C^2+\tfrac{4}{3}\Lambda D}) \in (-\infty,-1]\cup [1,\infty), \\
&\text{for } \Lambda>0: \ \  A^2-4B<0 \ \ \ \wedge \ \ \ C^2+\tfrac{4}{3}\Lambda D<0,
\end{align}
where the second column corresponds to the condition that $P^2>0$.  

We will separately investigate the case when the denominator of $\Omega$ (\ref{OmegaGen}) vanishes, namely
\begin{align}
    A^2=4B.
\end{align}
It follows from the definition of parametrization $(A,B,C,D)$ in (\ref{par1}) that this specific case corresponds to
\begin{align}\label{c2c1real}
   \text{Im} \Big[\frac{c_2}{c_1}\Big]=0.
\end{align}
Moreover, the parametrization degenerates from four to only two independent parameters and is not sufficient to cover the solution to the imaginary part of the Type D equation (\ref{typeDim}) - the rotation pseudoscalar $\Omega$. Consequently, one has to introduce a third parameter.
Notice, that whenever (\ref{c2c1real}) holds, the invariant $\Psi_2$ (\ref{psic1c2}) is of the form:
\begin{align}
    \Psi_2= \frac{1}{c_1^3(x+A')^3},
\end{align}
for some parameter $A'\in\mathbb{R}$. It is, then, convenient to introduce a pair of real parameters, $B'$ and $C'$, such that
\begin{align}
    B'&= \frac{1}{c_1^3}+\frac{1}{\bar c_1^3}, \ \ \ \ \ \ \ \ \ \ \ \ \ \ iC'= \frac{1}{c_1^3}-\frac{1}{\bar c_1^3}.
\end{align}
Consequently, the real (\ref{typeDexplicit}) and imaginary (\ref{typeDim}) parts of the Type D equation are of the form
\begin{align}
    \frac{B'}{2(x+A')^3} &= \frac{1}{4R^2}\partial_x^2 P^2 +\frac{\Lambda}{6},\\ 
   \frac{-C'}{2(x+A')^3} &=-\frac{1}{2}\Omega,
 \end{align}
and $A'\in(-\infty, -1)\cup (1,\infty)$. Their solutions read
\begin{align}\label{pspec}
    P^2&= R^2 \frac{B'-\frac{\Lambda}{3}(A'^2-1)(A'+x)}{(A'^2-1)(A'+x)}(x^2-1),\\ \label{omegaspec}
    \Omega &= \frac{C'}{(x+A')^3}.
\end{align}
Additionally, we have to restrict parameters $A'$ and $B'$, so that the frame coefficient is positive everywhere on the sphere (apart from the poles where it has to vanish), meaning the following must hold:
\begin{align}
    \frac{B'}{A'+x}< \frac{\Lambda}{3}\big( A'^2-1\big),
\end{align}
 for any $x\in(-1,1)$. 
 
 {\color{black} We go back now to the conditions (\ref{P^2poles}) - (\ref{Omcon}). The condition (\ref{P^2poles}) has been ensured from the beginning of our derivation.} 
 Remarkably, the conical singularity  condition (\ref{con}) is satisfied now automatically for both solutions parametrized by $(A,B,C,D)$ and $(A',B',C')$. So is the regularity condition (\ref{Omcon}), due to  the condition (\ref{P^2poles}) and the regularity of $\Omega$ as a function of $x$.
 {
\color{black} Finally, we can calculate the rescaling constants $\beta, \beta'$ (\ref{psi-phi}),
namely
 \begin{equation}
\begin{split}
    &\beta=\frac{-2}{\partial_x P^2}\eval_{x=1}=-\frac{1+A+B}{R^2 \left(C+D-\frac{\Lambda }{3}\right)},\\
    & \beta'=\frac{2}{\partial_x P^2}\eval_{x=-1}=\frac{1- A+B}{R^2 ( C- D+\frac{\Lambda}{3})}.
\end{split}
\end{equation}
Hence, our solutions {$(g_{AB},\Omega)$}, that is the pairs (\ref{Pfinal}), (\ref{OmegaGen}) and (\ref{pspec}), (\ref{omegaspec}), define uniquely a  vacuum isolated horizon structure of the Petrov Type D on the Hopf bundle $H$, that is continues at the points of the fibers over the poles of the sphere. However, it can be checked by inspection that the specific family of functions $P$ and $\Omega$ derived above make the horizon structure everywhere analytic \cite{Kaminski-private}.
The case when also $\beta=\beta'$ is satisfied has already been studied in \cite{hopf}.
}

\section{Embeddability of solutions to Petrov Type D equation}
\label{sec:embeddability}
\subsection{Accelerated Kerr-NUT-(anti-)de Sitter spacetimes} 
In this section we compare the nonembedded isolated horizons of the Hopf bundle $H\rightarrow S^2$, structure derived in the previous subsection as general solutions to the vacuum  Petrov Type D equation, with Killing horizons contained  in  known Petrov Type D spacetimes, exact solutions to the $\Lambda$-vacuum Einstein equations.  
Recently, spacetimes of the desired topology $H\times \mathbb{R}$ that are locally isometric to the accelerated Kerr-NUT-(anti-)de Sitter spacetimes were found \cite{LO3PhysRevD.104.024022}.
A general family of solutions to Einstein equations with cosmological constant of algebraic Type D has been first derived by Debever \cite{debever_1971} and later put in a more elegant and convenient form by Plebański and Demiański \cite{PLEBANSKI197698}. 
It contains generalized spherical black hole solutions with NUT parameter which may be described by the following metric tensor  obtained by Podolsky and Griffiths \cite{gp2006}. 
Setting the electric and magnetic charges to zero the metric tensor reads:
\begin{align}
\label{eq:acc-KNadS-metric}
ds^2=\frac{1}{F^2}\bigg[ -\frac{\mathcal{Q}}{\Sigma}(dt-Ad\phi)^2+\frac{\Sigma}{\mathcal{Q}}dr^2+\frac{\Sigma}{\mathcal{P}}d\theta^2+\frac{\mathcal{P}}{\Sigma}\sin^2\theta(adt-\rho d\phi)^2 \bigg]
\end{align}
where:
\begin{equation}
    \begin{split}
F&=1-\frac{\alpha}{\bar\omega}(l+a\cos\theta)r,\\
\Sigma&= r^2+(l+a\cos\theta)^2,\\
A&=a\sin^2\theta +4l\sin^2\frac{1}{2}\theta,\\
\rho&=r^2+(l+a)^2=\Sigma+aA,\\
\mathcal{Q}&=\bar \omega^2k-2Mr+\epsilon r^2-2\frac{\alpha \nu}{\bar\omega}r^3-(\alpha k +\frac{1}{3}\Lambda)r^4,\\
\mathcal{P}&=1-a_3\cos\theta-a_4\cos^2\theta,\\
a_3&=2\frac{\alpha a M}{\bar\omega} - \alpha^2alk-\frac{4}{3}\Lambda a l,\\
a_4&=-\alpha^2 a^2k-\frac{1}{3}\Lambda a^2,\\
\epsilon&=\frac{\bar\omega^2k}{a^2-l^2}+4\frac{\alpha lM}{\omega} - (a^2+3l^2)(\alpha^2k +\frac{1}{3}\Lambda),\\
\nu&=\frac{\bar\omega^2kl}{a^2-l^2}-\frac{\alpha M (a^2-l^2)}{\bar\omega}+(a^2-l^2)(\alpha^2 k +\frac{1}{3}\Lambda),\\
k&=\frac{1+2\alpha l M \bar \omega-l^2\Lambda}{3\alpha^2l^2+\bar \omega^2/(a^2-l^2)}.
    \end{split}
\end{equation}
Metric tensor (\ref{eq:acc-KNadS-metric}) is manifestly singular for such $r_H$ that $\mathcal{Q}(r_H)=0$. 
Therefore, similarly to a nonaccelerating case \cite{LO3PhysRevD.104.024022} we introduce Eddington-Finkelstein-like coframe transformation:
\begin{align}
dv:=dt+\frac{\rho}{\mathcal{Q}}dr && d\tilde \phi :=d\phi +\frac{a}{\mathcal{Q}}dr
\end{align}
which yields:
\begin{align}
\label{eq:metric-acc-knads-adv}
ds^2
&=\frac{1}{F^2}\bigg[ -\frac{\mathcal{Q}}{\Sigma}\big(dv-Ad\tilde \phi \big)^2+2\big(dv-Ad\tilde \phi\big)dr+\frac{\Sigma}{\mathcal{P}}d\theta^2+\frac{\mathcal{P}}{\Sigma}\sin^2\theta(adv-\rho d\tilde\phi)^2 \bigg].
\end{align}

The above metric depends on six parameters: mass parameter $M$, Kerr parameter $a$, NUT parameter $l$, acceleration parameter $\alpha$, cosmological constant $\Lambda$ and a gauge parameter $\bar\omega$ (note the change of name from the original $\omega$).
Thus, $\bar\omega$ is a free, nonphysical parameter and can be set to a convenient value to obtain various subclasses of the general solution by a limiting procedure.
Recently, a useful value has been proposed in \cite{PodolskyMatejov}:
\begin{equation*}
\bar\omega=\frac{a^2+l^2}{a}.
\end{equation*}
This value allows for the defining functions to be conveniently factorized; however, to obtain all subclasses a reparametrization is required.
We use the most widely spread form of the metric which keeps $\bar\omega$ free with full generality. 
Therefore, for a given cosmological constant it has four physical parameters: $M,\ a,\ l, \alpha$.
This motivates us to look for embedding of the four parameter family of isolated horizons defined in Sec \ref{sec:Generic solution for nonvanishing} in the accelerated Kerr-NUT-(anti-)de Sitter spacetimes.

In the metric tensor (\ref{eq:metric-acc-knads-adv}) there are several possible sources of singularities. 
The most obvious one is the vanishing of functions $\mathcal{P}$, $\Sigma$ and $F$.
We excluded vanishing of $\mathcal{P}$ simply by maintaining Lorentzian signature which requires that $\mathcal{P}>0$. Vanishing of $\Sigma=r^2+(l+a\cos\theta)^2$ corresponds to a ringlike, curvature singularity, similar to the singularity in the Kerr metric.
This may happen only when 
\begin{equation*}
    r=0 \text{ and } a^2>l^2.
\end{equation*}
Finally, the vanishing of $F$ corresponds to the conformal infinity.

From the point of view of the horizons more pressing are the next two possible singularities.
First, the 1-form $\hat\omega=dv-Ad\tilde{\phi}$ is not well-defined on the axis $\theta=\pi$.
Second, the metric on a (local) surface of constant $r$ and $v$ may not be a metric smoothly defined on $S^2$ and instead has a conical singularities at the poles $\theta=0$ and $\theta=\pi$.

The singularity of $\hat\omega$ can be solved by the procedure of so-called Misner's gluing and which amounts to introducing another coordinate $v'$ such that the chart containing $v'$ covers the southern pole smoothly.
Originally it was proposed by Misner \cite{misner} as a solution to the Taub-NUT singularity and it bears striking resemblance to Dirac's monopole construction.
It is a recent development that this procedure works also for the Kerr-NUT-(anti-)de Sitter and its accelerated counterparts, and has been described in the non-accelerated case in \cite{LO3PhysRevD.104.024022}.

In the nonaccelerating case $\alpha=0$ the conical singularity appears only when $a \neq 0$ \cite{LO3PhysRevD.104.024022}. Moreover, when the acceleration is present the nonvanishing NUT parameter is necessary for the conical singularity to vanish.
This follows from facts that C-metric as well as accelerated Kerr-(anti-)de Sitter have in irremovable singularity and that accelerated Taub-NUT is no longer of the Type D \cite{AcceleratingNUTblackholes}.
Thus, the only remaining possibility is the general accelerated Kerr-NUT-(anti-)de Sitter.
The conical singularity in at least one of the poles can always be remedied by changing the range of the angular coordinate.
This amounts to a rescaling by a parameter which we call $\beta$.
Yet this is not enough and in general the rescaling in the other pole would be different.
The solution is a procedure that is an extension of the Misner's interpretation of the Taub-NUT spacetime.

We now briefly state the construction of nonsingular spacetimes described in \cite{LO3PhysRevD.104.024022}. 
The original Misner's recipe is equivalent to compacting the orbits of the Killing vector field $\partial_t$ (or $\partial_v$ in coordinates covering the horizon).
This is equivalent to introducing a periodic time coordinate in standard Boyer-Lindquist-like coordinates (or a periodic null coordinate in Eddington-Finkelstein coordinates).
Unfortunately, this simple choice works only when $al\Lambda=0$.
A more general approach is required and the solution turns out to be allowing for the compactification of the orbits of an arbitrary Killing vector field. 
First, we choose an arbitrary Killing vector field being a constant linear combination of $\partial_v$ and $\partial_{\tilde{\phi}}$
\begin{equation}
\label{eq:glueing-killing}
\xi=\partial_v+b\partial_{\tilde{\phi}},
\end{equation}
where $b \in \mathbb{R}$.
The second step is a computation of the metric on the space of the orbits of $\xi$ and subsequently its pullback to the surface of $r=\text{const}$.
On this two-dimensional space we have a positive definite 2-metric possessing a conical singularity.

The condition for its removal is a regularity of the circles being integral curves of the rotation symmetry of the space of the orbits. 
Then, the following conditions have to be satisfied:
\begin{equation}
\label{eq:p0=pPigeneric}
P(0)=\frac{P(\pi)}{| 1 - 4 l b |}, \quad \beta=1/P(0).
\end{equation}
There are two branches of the solution:
\begin{align}
b_+=& \frac{a \left(3 \alpha ^3 l^2 m (l^2-a^2) +6 \alpha ^2 l \omega  (l^2-a^2) -2 \Lambda
l \omega ^3+3 \alpha  m \omega ^2\right)}{l \left(3 \alpha ^2 \omega  (l-a) (a+l)^2 (a+3 l)-\omega ^3
(a \Lambda  (a+4 l)+3)+6 a \alpha ^3 l m (l-a) (a+l)^2+6 a \alpha  m \omega ^2\right)},\\
b_-=& -\frac{\omega ^3 \left(a^2 \Lambda +3\right)+3 \alpha ^2 \omega  (a^2-l^2)
\left(a^2+3 l^2\right)+6 a^2 \alpha ^3 l m (a^2-l^2) }{2 l \left(3 \alpha ^2 \omega  (l-a) (a+l)^2
(a+3 l)-\omega ^3 (a \Lambda  (a+4 l)+3)+6 a \alpha ^3 l m (l-a) (a+l)^2+6 a \alpha  m \omega
^2\right)}.
\end{align}
However, it can be shown that it does not matter whether we choose $b_+$ or $b_-$, the constructed spacetime will be the same \cite{LO3PhysRevD.104.024022}.
From the point of view of the horizon it is more convenient to parametrize the solutions by $r_H$ instead of $M$.
Consequently, one has to solve the condition (\ref{eq:p0=pPigeneric}) simultaneously with $\mathcal{Q}=0$, which gives

\begin{align*}
b_+=& \bigg(a  (6 \alpha ^4 l r^3  (a^2-l^2 )^2-3 \alpha ^3 r^2 \omega  (a^2-l^2)  (a^2-5 l^2+r^2 )\\
&+2 \alpha ^2 l r \omega ^2  (a^2  (\Lambda  r^2-6 )+l^2
(6-2 \Lambda  r^2 )-3 r^2 )\\
&+\alpha  \omega ^3  (a^2 (\Lambda  (l-r) (l+r)+3)-3
l^2-\Lambda   (l^4-6 l^2 r^2+r^4 )+3 r^2 )-4 \Lambda  l r \omega ^4 )\bigg)\\
&\bigg(2 l
(\alpha  \omega ^3  (a \Lambda   (r^2  (-a^2+3 a l+6 l^2 )+l (a-l)
(a+l)^2-r^4 )+3 r^2 (a+3 l)+3 (a-l) (a+l)^2 )\\
&+\alpha ^2 r \omega ^2 (a+l)  (-3 a^3+a
\Lambda  r^2  (a^2+a l-4 l^2 )-9 a^2 l+3 a  (l^2+r^2 )+9 l (l^2-r^2) )\\
&+3\alpha ^4 r^3 (a-l)^2 (a+l)^4-3 \alpha ^3 r^2 \omega  (a-l) (a+l)^2  ((a-3 l) (a+l)+r^2 )\\
&-r
\omega ^4 (a \Lambda  (a+4 l)+3) )\bigg)^{-1},\\
M_+=& \bigg(\omega   (6 \alpha ^3 l r^3
(a^2-l^2 )^2+2 \alpha  l r^3 \omega ^2  (a^2 \Lambda -4 \Lambda  l^2+3 )+3 \alpha ^2
r^2 \omega  (a^2-l^2)  (a^2+3 l^2+r^2 )\\
&+\omega ^3  (a^2  (3 \Lambda  l^2+\Lambda 
r^2-3 )-3 \Lambda  l^4+l^2  (6 \Lambda  r^2+3 )+\Lambda  r^4-3 r^2 ) )\bigg) \\
&\bigg(6
(\alpha  l r-\omega ) (\alpha  r (a+l)-\omega ) (\alpha  r (a-l)+\omega ) (\alpha  l (a^2-l^2)-r
\omega )\bigg)^{-1},\\
b_-  =& \bigg(3 \alpha ^4 r^3  (a^2-l^2 )^2  (a^2+l^2 )+3 \alpha ^3 l r^2
\omega  (a^2-l^2) (a^2+3 l^2-r^2 )\\
&+\alpha  l \omega ^3  (a^2 \Lambda +3 )
(a^2-l^2+3 r^2 )-r \omega ^4  (a^2 \Lambda +3 )+\alpha ^2 r \omega ^2  (a^4 (\Lambda  r^2-3 )\\
&+3 a^2  (r^2-l^2  (\Lambda  r^2+2 ) )+9 l^2 (l^2-r^2))\bigg)\\
&\bigg(2 l  (\alpha  \omega ^3  (a \Lambda   (r^2  (-a^2+3 a l+6 l^2 )+l (a-l) (a+l)^2-r^4 )\\
&+3 r^2 (a+3 l)+3 (a-l) (a+l)^2 )+\alpha ^2 r \omega ^2 (a+l)  (-3
a^3+a \Lambda  r^2  (a^2+a l-4 l^2 )-9 a^2 l\\
&+3 a  (l^2+r^2 )+9 l (l-r) (l+r) )+3 \alpha ^4 r^3 (a-l)^2 (a+l)^4\\
&-3 \alpha ^3 r^2 \omega  (a-l) (a+l)^2  ((a-3 l) (a+l)+r^2 )-r
\omega ^4 (a \Lambda  (a+4 l)+3) )\bigg),\\
M_-=& \bigg(\omega   (6 \alpha ^3 l r^3
(a^2-l^2 )^2+2 \alpha  l r^3 \omega ^2  (a^2 \Lambda -4 \Lambda  l^2+3 )+3 \alpha ^2   r^2 \omega  (a-l) (a+l)  (a^2+3 l^2+r^2 )\\
&+\omega ^3  (a^2  (3 \Lambda  l^2+\Lambda
r^2-3 )-3 \Lambda  l^4+l^2  (6 \Lambda  r^2+3 )+\Lambda  r^4-3 r^2 ) )\bigg)\\
&\bigg(6
(\alpha  l r-\omega ) (\alpha  r (a+l)-\omega ) (\alpha  r (a-l)+\omega ) (\alpha  l (a-l) (a+l)-r
\omega )\bigg)^{-1}.
\end{align*}
Now for any choice of spacetime parameters $(\alpha,a,M,l)$ we may obtain $b$, which determines $\xi$.

Next we introduce adapted Killing coordinates
\begin{equation}
\label{eq:killingCoords}
(x^\mu)=(\tau,x^i) =   \big(v, \ r,\ \theta,\ {\hat\phi}:=\mathcal{P}(0)(-b v+ \Tilde{\phi})\big)
\end{equation}
in which the metric tensor (\ref{eq:metric-acc-knads-adv}) reads

\begin{equation}
\label{eq:metricThroughHorizonKillingCoords0}
\begin{split}
ds^2=&\frac{1}{F^2}\Bigg[-\frac{\mathcal{Q}}{\Sigma}\bigg(\Big(1-b A\Big)d\tau-\frac{A}{\mathcal{P}(0)} d{\hat\phi}\bigg)^2+2 dr \bigg(\Big(1-b A\Big)d\tau-\frac{A}{\mathcal{P}(0)} d{\hat\phi}\bigg)+\frac{\Sigma}{\mathcal{P}}d\theta^2+\\
&+\frac{\mathcal{P}}{\Sigma}\sin^2\theta\bigg(\Big(a-\rho b\Big) d\tau-\frac{\rho}{\mathcal{P}(0)} d{\hat\phi}\bigg)^2\Bigg].
\end{split}
\end{equation}
The final step is performing the Misner gluing 
\begin{equation}
\label{eq:t'}
\tau=\tau' + \sigma\frac{4l}{P(\pi)}{\hat\phi}' , \ \ \ r=r', \ \ \ \theta=\theta', \ \ \ {\hat\phi}={\hat\phi}'\ \ \ \ \ {\rm for} \ \ \ \theta,\theta'\not= 0,\pi,
\end{equation}

\begin{align}
ds'^2&=\frac{1}{F^2}\Bigg[ -\frac{\mathcal{Q}}{\Sigma}\bigg(\Big(1-bA\Big)d\tau'-\frac{A'}{ \mathcal{P}(\pi)}\sigma d\hat\phi' \bigg)^2+2\bigg(\Big(1-bA\Big)d\tau'-\frac{A'}{ \mathcal{P}(\pi)}\sigma d\hat\phi' \bigg)dr'+\frac{\Sigma}{\mathcal{P}}d\theta'^2\nonumber\\
&+\frac{\mathcal{P}}{\Sigma}\sin^2\theta\bigg(\Big(a-\rho b\Big)d\tau'-\frac{\rho'}{ \mathcal{P}(\pi)}\sigma d\hat\phi' \bigg)^2 \Bigg],
\end{align}
where 
\begin{equation*}
\rho'=r^2+(a-l)^2\quad A' = a\sin^2\theta'-4l\cos^2(\theta/2)
\end{equation*}
and determining the range of $\tau$ to be 
\begin{equation}\label{tau_0}
\tau_0=2\pi \frac{4l}{1-4l b}\frac{1}{P(0)}.
\end{equation}

For the special case of $\xi$ generating the horizon, i.e. $b=\frac{a}{\rho_H}$, removing the conical singularity on the space of the orbits  of $\xi$ simultaneously removes the conical singularity on the space of the null generators of the Killing horizon of radius $r_H$.
Such horizons have been called \textit{projectively nonsingular} (i.e. possessing a nonsingular space of the null generators) and have been studied in the case of the Kerr-NUT-de Sitter metrics \cite{LO2}. However, in the generic case the space of the null generators may still suffer from conical singularity, even if the ambient space is smooth. 

In this paper we focus on the case when $\xi$ is not null and seek to embed the solutions to Petrov Type D equation found in Sec \ref{sec:Generic solution for nonvanishing} as such horizons. The metrics obtained above are well-defined for vanishing $\mathcal{Q}$, and now one may consider null surfaces characterized by $r=r_H$. Their metrics are degenerate and read
\begin{align}
\label{eq:metric-3d-horizon-b}
ds_{\mathcal{H}}^2&=\frac{1}{F^2}\Bigg[ \frac{\Sigma_H}{\mathcal{P}}d\theta^2+\frac{\mathcal{P}}{\Sigma_H}\sin^2\theta\bigg(\Big(a-\rho_H b\Big)d\tau-\frac{\rho_H}{ \mathcal{P}(0)} d\hat\phi \bigg)^2 \Bigg],
\end{align}
and
\begin{align}
{ds'}_{\mathcal{H'}}^2 &=\frac{1}{F^2}\Bigg[\frac{\Sigma_H}{\mathcal{P}}d{\theta'}^2+\frac{\mathcal{P}}{\Sigma_H}\sin^2\theta\bigg(\Big(a-\rho_H b\Big)d\tau'-\frac{\rho'_H}{ \mathcal{P}(\pi)}\sigma d\hat\phi' \bigg)^2 \Bigg].
\end{align}
Now to descend to the space of null generators we introduce new coordinates $\varphi$ and $\varphi'$,
\begin{align}
\label{eq:coords-horizon-def}
\varphi = \hat\phi - \frac{\mathcal{P}(0)}{\rho_H}(a-\rho_H b)\tau, && \varphi' = \hat\phi' - \frac{\mathcal{P}(\pi)}{\rho'_H}(a-\rho_H b)\sigma \tau',
\end{align}
which yield
\begin{align}
ds_{\mathcal{H}}^2&=\frac{1}{F^2}\Bigg[ \frac{\Sigma_H}{\mathcal{P}}d\theta^2+\frac{\mathcal{P}\rho_H^2}{\mathcal{P}^2(0)\Sigma_H}\sin^2\theta d \varphi^2 \Bigg]
\end{align}
and
\begin{align}
{ds'}_{\mathcal{H'}}^2 &=\frac{1}{F^2}\Bigg[\frac{\Sigma_H}{\mathcal{P}}d{\theta'}^2+\frac{\mathcal{P}{\rho'}_{H}^{2}}{\mathcal{P}^2(\pi)\Sigma_H}\sin^2\theta \sigma   d\varphi'^2 \Bigg].
\end{align}
Notice that the above procedure is an analog of the horizon gluing presented in Sec \ref{horizongluing}. 
Moreover, the rotation 1-form potential $\omega$ is well-defined as it satisfies conditions (\ref{omega_0}) and (\ref{omega_pi}) on the poles. 
On the other hand, the conditions for a well-defined 1-form $\omega$ provide the correct spacetime gluing introduced in (\ref{eq:t'}).

This can be seen in two ways.
First the 1-form $\omega$ can be calculated and expressed in coordinates $(\tau,\theta,\varphi)$ and $(\tau',\theta',\varphi')$ adopted to the space of null generators.
Since both $\omega$ and $\kappa$ depend on the normalization of $\ell$ let us first make the obvious choice of (expressed in the advanced coordinates (\ref{eq:metric-acc-knads-adv}))
\begin{equation}
    \ell=\partial_v+\frac{a}{\rho_H}\partial_{\tilde\phi}.
\end{equation}
Then, $\kappa$ reads simply
\begin{equation}
    \kappa=\frac{1}{2}\frac{\mathcal{Q}'(r)\eval_{r=r_H}}{\rho_H},
\end{equation}
and $\omega$ is of the form
\begin{equation}
    \omega=\kappa d v+\omega_\theta d \theta+\frac{\omega_{\phi_H}}{\mathcal{P}(0)} d\varphi=\kappa\frac{\rho'_H}{\rho_H}(1-4lb) d v'+\omega_\theta d \theta'+\sigma\left(4l\kappa +\omega_{\phi_H}\frac{\rho'_H}{\rho_H}\right)\frac{1}{\mathcal{P}(\pi)} d\varphi',
\end{equation}
where:
\begin{equation}
    \omega_\theta=-\frac{2\Sigma \partial_\theta F+aF\partial_\theta A}{2 F \Sigma}\eval_{r=r_H}=a\sin\theta\frac{\frac{r_H^3\alpha}{\bar\omega}+l+a\cos{\theta}}{F_H\Sigma_H}
\end{equation}
\begin{equation}
    \omega_{\phi_H} =-\frac{ a \rho r \mathcal{P}\sin^2\theta+Q'A \Sigma}{ \Sigma^2}\eval_{r=r_H}=-\frac{ a \rho_H r_H \mathcal{P}\sin^2\theta}{\Sigma_H^2}-\kappa A \frac{ \rho_H}{\Sigma_H},
\end{equation}
making $\omega$ explicitly regular at $\theta=0$ and $\theta=\pi$ in unprimed and primed coordinates, respectively.

In the above we used the fact that orbits of Killing vector field (\ref{eq:glueing-killing}) with the constant $b$ satisfying condition (\ref{eq:p0=pPigeneric}) have been compactified and the axis of rotation is regular in spacetime.
However, condition (\ref{eq:p0=pPigeneric}) can be recovered purely from the horizon data by the procedure described in the Sec. \ref{horizongluing}.
The starting point is the degenerate metric tensor on the horizon (\ref{eq:metric-3d-horizon-b}) adopted to the Killing vector field (\ref{eq:killingCoords}) generating the action transversal to the horizon.
The first step is putting (\ref{eq:metric-3d-horizon-b}) into the form of (\ref{Hmetric}).
This is achieved by taking
\begin{equation}
\label{eq:functions-transversal-construction-aknads}
    U^2(\theta)=\frac{\Sigma_H}{\mathcal{P}F^2},\quad f(\theta)=\frac{\mathcal{P}\rho_H}{\Sigma_H},\quad f(0)=\mathcal{P}(0),\quad f(\pi)=\mathcal{P}\frac{\rho_H}{\rho'_H},
\end{equation}
and defining $v= \gamma \tau$, where:
\begin{equation*}
    \gamma=\frac{a-\rho_H b}{\rho_H}\left(\frac{1}{f(0)}-\frac{1}{f(\pi)}\right)^{-1}.
\end{equation*}

The second step is to normalize $\ell$ so that having chosen (\ref{eq:functions-transversal-construction-aknads}) the vector fields are of the same form as (\ref{eq:ell-transversal-construction}).
To do so we rescale
\begin{equation}
    \ell=\frac{f(\pi)}{\gamma}\left(\partial_v+\frac{a}{\rho_H}\partial_{\tilde\phi}\right).
\end{equation}
With the above normalization the rotation 1-form $\omega$ can be calculated and put in the form (\ref{omegaFinalpi}) with 
\begin{equation}
    h(\theta)=\frac{\omega_{\phi_H} \gamma}{f(\pi)}+\frac{\kappa f(0)}{f(0)+f(\pi)}.
\end{equation}
Then, solving condition (\ref{ConditionKappa}) precisely reproduces relation (\ref{eq:p0=pPigeneric}) between the Killing vector field parameter $b$ and the spacetime parameters.

Therefore, not only the number of parameters but also the structure of the horizons motivates us to investigate the embeddability of the generic solution in the accelerated Kerr-NUT-(anti-)de Sitter spacetime, which will be a subject of the next section.
For simplicity, we express both of the above metrics by one expression:
\begin{align}
g_{ab}dx^adx^b=\frac{1}{F^2}\bigg( \frac{\Sigma_H}{\mathcal{P}}d\theta^2+\frac{\mathcal{P}}{\Sigma_H}\rho_H^2\sin^2\theta \beta^2d\varphi^2\bigg),
\end{align}
where $\beta$ is of a different form depending on what chart we refer to, that is:
\begin{align}
    \beta_\mathcal{H} = \frac{1}{\mathcal{P}(0)} &&\text{and}&& \beta_\mathcal{H'} = \frac{\rho'_H}{\mathcal{P}(\pi)\rho_H}.
\end{align}

\subsection{Comparison of the 2D geometries}
\label{sec:Embeddability}
We now investigate embeddability of the generic solution to the Type D equation in the accelerated Kerr-NUT-(anti-)de Sitter spacetime. 
We begin with the comparison of the two metrics of two-dimensional surfaces:
\begin{align}
R^2\bigg( \frac{1}{P^2}dx^2+ P^2 {\color{black} \beta^2}d\varphi^2 \bigg)=\frac{1}{F^2}\bigg( \frac{\Sigma_H}{\mathcal{P}}d\theta^2+\frac{\mathcal{P}}{\Sigma_H}\rho_H^2\sin^2\theta \beta^2d\varphi^2\bigg).
\end{align}
 Notice, that the metric on the left-hand side above, satisfying the Type D equation, is of the general form (\ref{genMetric}), {\color{black} with the coordinate $\varphi\in[0,2\pi[$ introduced as in (\ref{psi-phi})}, which allows for a construction of the isolated horizon $H$ diffeomorphic to $S^3$, just as the accelerated Kerr-NUT-(anti-)de Sitter horizon.

From the above comparison we find the expression for the area radius $R$ written in terms of the accelerated Kerr-NUT-(anti-)de Sitter parameters, namely
\begin{align}
\label{eq:R2}
R^2=\frac{\rho_H\bar \omega^2}{(\bar \omega+r_H\alpha(a-l))(\bar \omega-r_H\alpha(a+l))}=:\rho_H \xi.
\end{align}
Moreover, the frame coefficient takes the form
\begin{align}
P^2&=\frac{\mathcal{P}}{F^2\Sigma_H\xi}\rho_H\sin^2\theta .
\end{align}
Next, we calculate the relation between the coordinates $x$ and $\theta$:
\begin{align}\label{RKNUT}
R^2 \frac{1}{P^2}dx^2&=\frac{1}{F^2} \frac{\Sigma_H}{\mathcal{P}}d\theta^2\\
dx&=\frac{1}{F^2\xi}\sin\theta d\theta
\end{align}
which yields
\begin{align}\label{x=cos}
x = \frac{\cos\theta(\bar\omega-\alpha lr_H)-a\alpha r_H}{a\alpha r_H\cos\theta+\alpha l r_H-\bar\omega},
\end{align}
or equivalently
\begin{align}\label{cos=x}
\cos\theta=\frac{a\alpha r_H+(\alpha lr-\bar \omega)x}{\bar \omega-\alpha lr_H-a \alpha r_H x}.
\end{align}
Therefore, the frame coefficient written explicitly in terms of $\theta$ reads
\begin{align}
P^2&=\frac{\mathcal{P}}{F^2\Sigma_H\xi}\rho_H\sin^2\theta \nonumber\\
&=\frac{1-a_3\cos\theta-a_4\cos^2\theta}{(1-\frac{\alpha}{\bar \omega}(l+a\cos\theta)r_H)^2( r_H^2+(l+a\cos\theta)^2)\xi}\rho_H\sin^2\theta .
\end{align}
On the other hand, we found the general solution (\ref{Pfinal}) to the real part of the Type D equation (\ref{typeDexplicit}), which together with (\ref{RKNUT}) yields
\begin{align}\label{Pgeneric}
P^2&=\frac{\beta\bar \omega^2\rho_H(-\frac{\Lambda}{3}x^2+Cx+D)}{(x^2+Ax+B)\big((a-l)r\alpha+\bar \omega \big)\big( -(a+l)r\alpha+\bar \omega\big)}(x^2-1).
\end{align}
Comparing the two expressions for $P^2$ gives
\begin{align}
\frac{(-\frac{\Lambda}{3}x^2+Cx+D)(x^2-1)}{(x^2+Ax+B)}=\frac{(1-a_3\cos\theta-a_4\cos^2\theta){(\bar \omega+r_H\alpha(a-l))^2(\bar \omega-r_H\alpha(a+l))^2}}{(1-\frac{\alpha}{\bar \omega}(l+a\cos\theta)r_H)^2( r_H^2+(l+a\cos\theta)^2)\bar \omega^4}\sin^2\theta.
\end{align}
The above expression together with (\ref{x=cos}) [or (\ref{cos=x})] provides the relation between parameters $(A,B,C,D)$ and $(r_H, \alpha, l, a)$:
\begin{align}
A&=\frac{2 l r_H^4 \alpha^2 - 2 r_H\alpha\bar\omega (a^2 - l^2 + r_H^2)  - 2 l \bar \omega^2}{a (r_H^4 \alpha^2 + \bar \omega^2)},\\
B&=\frac{a^4 r_H^2 \alpha^2 - 2 a^2 l r_H \alpha (l r_H \alpha - \bar \omega) + (l^2 +  r_H^2) (l r_H \alpha - \bar \omega)^2}{a^2 (r_H^4 \alpha^2 + \bar \omega^2)},\\
C&=\frac{1}{ a \bar \omega (l r_H \alpha -\bar \omega) ((a^2 - l^2) l \alpha - r_H \bar \omega) (r_H^4 \alpha^2 + \bar \omega^2)}\bigg[- r_H^4\alpha^5 (a^2 - l^2)^2 (a^2 - l^2 - r_H^2) + \tfrac{4}{3} l r_H {\Lambda} \bar \omega^5\nonumber\\
&+ r_H^2 \alpha^3\bar \omega^2 \Big(2 (a^2 - 3 l^2) (a^2 - l^2 - r_H^2)+ \big((l^3-a^2 l )^2 - (a^4 - 11 a^2 l^2 + 14 l^4) r_H^2 + (a^2+ l^2) r_H^4\big) \tfrac{\Lambda}{3}\Big) \nonumber\\  
& + \alpha \Big( (l^2 +  r_H^2) + (l^4 - 14 l^2 r_H^2 + r_H^4) \tfrac{\Lambda}{3} -a^2 (1 + l^2 \tfrac{\Lambda}{3} - r_H^2 \Lambda)\Big)\bar \omega^4  -4 l  r_H^3 \alpha^4\bar \omega (a^2 - l^2)\nonumber\\
&\Big( a^2 -  r_H^2 +  l^2 ( r_H^2\tfrac{ \Lambda}{3}-1)\Big) \nonumber\\
& -2 l r_H \alpha^2 \Big(2 (-a^2 + l^2 + r_H^2) + \big((a^2 - l^2)^2 + 2 (2 a^2 - 5 l^2) r_H^2 + r_H^4\big) \tfrac{\Lambda}{3}\Big) \bar \omega^3\bigg],\\
D&=\frac{-1}{ a^2 \bar \omega^2 (l r_H \alpha - \bar \omega) ((a^2 - l^2) l \alpha  - r_H \bar \omega) (r_H^4 \alpha^2 + \bar \omega^2)} \bigg[ (a^2 - l^2)^4 r_H^5 \alpha^6 \nonumber\\
&+   l  r_H^4\alpha^5(a^2 - l^2)^2 (5 a^2 - 5 l^2 + r_H^2)\bar \omega  \nonumber\\
&-r_H^3 \alpha^4  (a^2 - l^2) \Big((a^2 - 5 l^2) (2 a^2 - 2 l^2 + r_H^2) -  a^2 (a^2 - 4 l^2) r_H^2 \tfrac{\Lambda}{3}\Big) \bar \omega^2 \nonumber\\
&- l r_H^2 \alpha^3 \Big(2 (3 a^2 - 5 l^2) (a^2 - l^2 + r_H^2) - a^2 \big((a^2 - l^2)^2 + 2 (3 a^2 - 5 l^2) r_H^2 + r_H^4\big) \tfrac{\Lambda}{3}\Big) \bar \omega^3 \nonumber\\
&+ r_H \alpha^2 \Big( (a^2 - 5 l^2) (a^2 - l^2 + 2 r_H^2) + a^2 \big(-l^4 + 10 l^2 r_H^2 - r_H^4 +  a^2 (l^2 - 2 r_H^2)\big) \tfrac{\Lambda}{3}\Big) \bar \omega^4 \nonumber\\
&- l \alpha \Big( (l^2 - 5 r_H^2) - a^2 (1 - \tfrac{4}{3} r_H^2 \Lambda)\Big) \bar \omega^5 - r_H \bar \omega^6\bigg].
\end{align}
Using the above relations we retrieve the expression for rotation pseudoscalar $\Omega$ for the accelerated Kerr-NUT-(anti)-de Sitter spacetime:
\begin{align}\label{OmegaGeneric}
    \Omega = \text{Im}\bigg[ \frac{(-iM+\nu)\Big(a^2\alpha^2r_H^2-(l\alpha r_H-\omega)^2\Big)^3}{\omega^3(-a^2\alpha r_H+(l-ir_H)(l\alpha r_H-\omega)-ax(i\alpha r_H^2-\omega))^3} \bigg].
\end{align}

Furthermore, we evaluate the determinant of the Jacobi matrix:
\begin{align}
J&=8 r_H^2  \bigg(\big( \bar \omega-l r_H \alpha \big)^2-a^2 r_H^2 \alpha^2\bigg)^6 \bigg( \big(a^2 - l^2\big)^2  \big(a^2 - l^2 - r_H^2\big) \alpha^4r_H^4 \\
&-4\big(a^2-l^2\big)  \big(-a^2 + l^2 + r_H^2\big) l\alpha^3 \bar \omega r_H^3 \nonumber + r_H^2 \alpha^2 \Big(a^4 \big(-2 +  l^2 \Lambda + \tfrac{1}{3}r_H^2 \Lambda \big) \\
&+ l^2 \big(l^2 + r_H^2\big)\big (-6 + (l^2 + r_H^2) \Lambda\big) \\ 
&+a^2\big (2 (4 l^2 + r_H^2) - (2 l^4 +  l^2 r_H^2 +\tfrac{1}{3} r_H^4) \Lambda \big)\Big) \bar \omega^2 \nonumber\\
&- 2 l r_H \alpha \Big(\tfrac{1}{3} a^4 \Lambda + a^2 (2 - \tfrac{2}{3} l^2 \Lambda) +  \big(l^2 +  r_H^2\big) \big(-2 + (l^2 + r_H^2) \Lambda\big)\Big) \bar \omega^3 \nonumber\\
&+\Big (a^2 (1 -  l^2 \Lambda + \tfrac{1}{3}r_H^2 \Lambda) + (l^2 + r_H^2) (-1 + (l^2 + r_H^2) \Lambda)\Big) \bar \omega^4\bigg)\nonumber\\
& \bigg/ \bigg( a^7 \bar \omega^3 (-l r_H \alpha + \bar \omega)^2 ( ( l^2-a^2)l \alpha + r_H \bar \omega)^2 (r_H^4 \alpha^2 + \bar \omega^2)^5\bigg).
\end{align}
It approaches infinity in the following cases:
\begin{align}
&J\rightarrow \infty \ \ \text{for} \  \ a= 0 \ \ \vee \ \bar \omega= 0 \ \vee \ \bar \omega=lr_H \alpha \ \vee \ \bar \omega = \tfrac{(a^2-l^2)l\alpha}{r_H}  
\end{align}
which coincide with the values for which parameters $A$, $B$, $C$, $D$ are ill-defined. Next, we consider the case for which the determinant $J$ of the Jacobi matrix vanishes:
\begin{align}\label{eq:J=0a}
&J=0 \ \text{for} \ \ r_H = 0 \ \vee \ r_H = \tfrac{\bar \omega}{(l\pm a)\alpha}, 
\end{align}
or whenever
\begin{align}
\label{eq:J=0b}
\big(&a^2 - l^2\big)^2  \big(a^2 - l^2 - r_H^2\big) \alpha^4r_H^4 - 4\big(a^2-l^2\big)  \big(-a^2 + l^2 + r_H^2\big) l\alpha^3  r_H^3\bar \omega \nonumber\\
& + r_H^2 \alpha^2 \Big(a^4 \big(-2 +  l^2 \Lambda + \tfrac{1}{3}r_H^2 \Lambda \big) + l^2 \big(l^2 + r_H^2\big)\big (-6 + (l^2 + r_H^2) \Lambda\big) \\
&+a^2\big (2 (4 l^2 + r_H^2) - (2 l^4 +  l^2 r_H^2 +\tfrac{1}{3} r_H^4) \Lambda \big)\Big)\bar \omega^2 \nonumber\\
&- 2 l r_H \alpha \Big(\tfrac{1}{3} a^4 \Lambda + a^2 (2 - \tfrac{2}{3} l^2 \Lambda) +  \big(l^2 +  r_H^2\big) \big(-2 + (l^2 + r_H^2) \Lambda\big)\Big) \bar \omega^3 \nonumber\\
&+\Big (a^2 (1 -  l^2 \Lambda + \tfrac{1}{3}r_H^2 \Lambda) + (l^2 + r_H^2) (-1 + (l^2 + r_H^2) \Lambda)\Big) \bar \omega^4=0.
\end{align}

The first part of the alternative in (\ref{eq:J=0a}) requiring $r_H=0$ further constrains the parameters since the zeroth coefficient of the polynomial $\mathcal{Q}$ has to vanish.
Solving this constraint we get
\begin{equation}
    \alpha=\frac{\omega(l^2\Lambda-1)}{2 l M}.
\end{equation}
Then, the map between parameter spaces degenerates to
\begin{equation}
    \begin{split}
        &A=-\frac{2l}{a},\\
        &B=\frac{l^2}{a^2},\\
        &C=\frac{\Lambda l^2 + 3}{3 a l},\\
        &D=-\frac{1}{a^2},
    \end{split}
\end{equation}
However, the above implies that $A^2-4B=0$, which corresponds to the case when rotation pseudoscalar $\Omega$ (\ref{OmegaGen}) is not well-defined and the alternative parameterization, $(A',B',C')$, is applicable.
Comparing the 2-metrics on the slices of the isolated horizon and accelerated Kerr-NUT-(anti-)de Sitter horizon with $r_H=0$, provides the following relation between our parameters $A'$, $B'$ and $a$, $l$:
\begin{align}
    A'&=-\frac{l}{a},\\
    B'&=\frac{l^2-a^2}{la^3}\bigg(1-\frac{\Lambda}{3}l^2\bigg).
\end{align}
Next, we calculate the rotation scalar $\Omega$ for the vanishing $r_H$ and obtain
\begin{align}\label{omegarh0}
    \Omega = \frac{M}{(l-ax)^3}.
\end{align}
It is, then, straightforward to find the correspondence between parameters $M$ and $C'$ using the expressions (\ref{omegarh0}) and
(\ref{omegaspec}). The result yields
\begin{align}
    C' =- \frac{M}{a^3}.
\end{align}

It is noteworthy that horizons at $r_H=0$ enjoy the benefit of a simplified form.
The conformal factor reads simply $F=1$ and consequently $x=-\cos\theta$.
The horizon radius can be expressed as $R^2=\beta (a^2+l^2)$ and the frame coefficient  as
\begin{equation}
    P^2= (a^2+l^2)\frac{\left(x^2-1\right) (a \Lambda  l x-3)}{3 l (l-a x)}.
\end{equation}
The smoothness conditions (\ref{beta}) and (\ref{eq:Lambda-smooth}) (soon to be introduced for a general case) read
\begin{equation}
    \beta=\frac{l^2}{a^2+l^2},\quad \Lambda=\frac{3}{l^2},
\end{equation}
which inserted in the above expression for $P^2$ and $R^2$ give
\begin{equation}
    P^2=1-x^2, \quad R^2=\frac{3}{\Lambda}.
\end{equation}
Those are precisely the defining properties of the Class II of solutions to the Petrov Type D equation on the horizon with a structure of a Hopf bundle and the $U(1)$ action coinciding with the generator of the horizon given in \cite{hopf}.


The second part of (\ref{eq:J=0a}), i.e. the condition that $r_H=\frac{\bar\omega}{(l\pm a)\alpha}$ corresponds to an acceleration horizon.
It can be seen by examining the conformal factor $F$, which now reads
\begin{equation}
    F=1-\frac{l+a\cos\theta}{l+a},
\end{equation}
and vanishes for $\theta=0$.
This implies that the horizon intersects Scri at the surface of $\theta=0$.
Moreover, our parametrization explicitly excludes this case as $R^2$ defined by (\ref{eq:R2}) becomes infinite.

The last equation (\ref{eq:J=0b}) is equivalent to the following condition for the cosmological constant
\begin{align}\label{lambdaJ}
\Lambda =& 3\big(a^2-l^2-r_H^2\big) \Big(\big(\bar \omega-\alpha l r_H \big)^2 -a^2\alpha^2r_H^2\Big)^2 \bigg/\\
&\Bigg(\bar \omega^2\bigg( -3\big(l^2+r_H^2 \big)^2 \big( \bar \omega-\alpha l r_H \big)^2\nonumber-a^4\alpha  r_H(3\alpha l^2 r_H +\alpha r_H^3-2 l \bar \omega)\\
&+a^2 \Big( 6\alpha^2 l^4 r_H^2 +\alpha^2 r_H^6 -8\alpha l^3 \bar \omega r_H - \bar \omega^2 r_H^2+3l^2(\bar \omega^2+\alpha^2 r_H^4) \Big) \bigg)\Bigg).
\end{align}
Interestingly, the above expression for $\Lambda$  is precisely the condition for the horizon to be extremal, therefore, the relation is locally invertible if and only if the horizon is nonextremal. 

Taking all into consideration, we conclude that the relation between the metric on a slice of isolated horizon and horizons of accelerated Kerr-NUT-(anti-)de Sitter is locally invertible for parameters corresponding to the generic case when four distinct horizons exist.

We have determined the explicit map between the spacetime parameters and the parameters $A,B,C,D$ (or $A',B',C'$) in one direction.
However, in the neighborhood of a bifurcated Type D horizon a unique Type D spacetime may be constructed \cite{Cole_2018}.
All four-dimensional $\Lambda-$vacuum Type D solutions to Einstein equations are contained in Plebański-Demiański family and it is our suspicion that the spacetimes constructed in the neighborhood of the spherical horizon would also be a spherical accelerated Kerr-NUT-(anti-)de Sitter.
Then for such a spacetime one may construct the spacetime admitting the bundle structure, containing the nonbifurcated horizon of the Hopf bundle structure.
If that would be the case the relation between the parameters would be a bijection, up to parameters defining the same spacetime.

\subsection{Differentiability condition for the case of the fibers coinciding with the null curves}\label{sec:Differentiability condition}
In the case of the horizon of a Hopf bundle structure such that the null vector field $\ell$ is tangent to the fibers of the bundle, requiring the metric (\ref{2metric}) to be simultaneously differentiable at both poles yields:
\begin{align}\label{diff}
\partial_xP^2|_{x=\pm1} = \mp2.
\end{align}
Differentiability at one pole may be obtained by finding the appropriate value for the parameter $\beta$, and for the pole $x=-1$ it is of the form
\begin{align}\label{beta}
    \beta=&\frac{1}{\mathcal{P}(0)}=\bigg((l r_H \alpha - \bar\omega) (( l-a) r_H \alpha - \bar\omega) ((a + l) r_H \alpha - \bar\omega) \big( (l^2-a^2) l\alpha + r_H \bar\omega\big)\bigg)\nonumber \\
    &\bigg / \bigg( (a - l)^2 (a + l)^4 r_H^3 \alpha^4 -  (a - l) (a + l)^2 ((a - 3 l) (a + l)  + 
    r_H^2)r_H^2  \alpha^3 \bar\omega \nonumber\\
    &+(a + l) (- a^3 - 
    3 a^2 l + 3 l (l - r_H) (l + r_H) +  a (l^2 + r_H^2) + 
     (a^2 + a l - 
       4 l^2)a r_H^2 \tfrac{\Lambda}{3}) r_H \alpha^2\bar\omega^2 \nonumber\\
       &+ ( (a - 
       l) (a + l)^2 +  (a + 3 l) r_H^2 + 
    a ((a - l) l (a + l)^2 + (-a^2 + 3 a l + 6 l^2) r_H^2 - 
       r_H^4) \tfrac{\Lambda}{3})\alpha  \bar\omega^3\nonumber\\
       &- 
  (1 + a (a + 4 l) \tfrac{\Lambda}{3}) r_H\bar\omega^4 \bigg).
\end{align}
There are, however, various special cases for the metric to be automatically differentiable at the other pole, such as:
\begin{align}\label{diff1}
a&=0 \ \ (\Rightarrow\alpha=0),\\\label{diff2}
l&=0 \  \wedge \ \alpha=0, \\\label{diff3}
\alpha&=0 \ \wedge \  \Lambda= \frac{3}{a^2+2l^2+2r^2_H}.
\end{align}
Another special case is $\Lambda=0$, which implies that the acceleration parameter $\alpha$ has to satisfy one of the following qualities:
\begin{align}\label{alpha1}
\alpha&=\frac{(a^2 - l^2 + r_H^2)^2 \bar \omega - \sqrt{\big(16 (a^2 - l^2) l^2  r_H^4 + (a^2 -l^2 + r_H^2)^4\big) \bar \omega^2}}{4 (a^2 - l^2) l r_H^3},\\\label{alpha2}
\alpha&=\frac{(a^2 - l^2 + r_H^2)^2 \bar \omega +\sqrt{\big(16 (a - l) l^2 (a + l) r_H^4 + (a^2 - l^2 +  r_H^2)^4\big) \bar \omega^2}}{4  (a^2 - l^2) lr_H^3}.
\end{align} 
In a generic case ($a,l\neq0$), requiring differentiability at both poles yields
\begin{align}
\label{eq:Lambda-smooth}
\Lambda&=-\bigg(3 \Big((a + l) r_H \alpha - \bar \omega\Big) \Big((a -  l) r_H \alpha + \bar \omega\Big) \Big(2 l (l^2-a^2)  r_H^3 \alpha^2 + (a^2 - l^2 +  r_H^2)^2 \alpha \bar \omega + 2 l r_H \bar \omega^2\Big)\bigg)\nonumber\\
&\Big/\bigg(\bar \omega^2 \Big(2 l r_H^3 \big(2 l^2 (l^2 + r_H^2) - a^2 (2 l^2 + r_H^2)\big) \alpha^2 + (l^2 + r_H^2) (a^2 - l^2 - 2 l r_H + r_H^2)\\
&(a^2 - l^2 + 2 l r_H +  r_H^2) \alpha \bar \omega \nonumber+ 2 l r_H \big(a^2 + 2 (l^2 + r_H^2)\big) \bar \omega^2\Big)\bigg)
\end{align}
which for $\alpha=0$ reduces to the known condition for nonaccelerating black holes \cite{LO}:
\begin{align}
\label{eq:Lambda-smooth-knads}
\Lambda= \frac{3}{a^2+2l^2+2r^2_H}.
\end{align}

\section{Summary}   

In this work we consider three-dimensional isolated horizons (i.e. potentially embeddable in four-dimensional spacetime) that admit the Hopf bundle structure. 
A new element is allowing for cases when the null direction is not tangent to the bundle fibers. 
In order to characterize them in a coordinate independent way, we identify the three-dimensional manifold, on which the horizon structures are considered, with the manifold of the Lie group SU(2).
The right action of the subgroup U(1) provides the Hopf bundle structure. 
The left action of U(1) defines rotations of the bundle, i.e. the circular rotational symmetry. 
The null vector field defining the canonical null symmetry of the horizon is assumed to be an arbitrary linear combination of the generators of the left and right actions of U(1).
The horizon structure is assumed to be invariant with respect to the two-dimensional group generated by the right and left actions of U(1). 
The surface gravity is assumed to be constant (the zeroth law of black hole thermodynamics); however, Einstein equations are not assumed at that point.    

Our first result is the characterization of the geometry of such horizons (we focus on the degenerate metric tensor and the rotation 1-form potential) by spherically symmetric data defined on a (topological) two-dimensional sphere: a metric tensor defined up to a constant rescaling with conical singularities at the poles, and a function regular at the poles - the rotation scalar. 
The surface gravity is determined by that data. 
If it is zero, then one more function is needed to reconstruct the rotation 1-form potential. 
Given any data characterized above, we construct a degenerate metric tensor and rotation potential 1-form defined regularly on the Hopf bundle.
This kind of data was considered in the literature as singular solutions to horizon equations, either extremal vacuum horizon \cite{Lucietti2013} or the Petrov Type D vacuum horizon \cite{Racz_2007}. 
Our result allows one to turn the singular solutions into nonsingular ones by replacing the $S^2\times \mathbb{R}$ topology of horizons by the $S^3$ topology, provided the singularity of the metric tensor originally given on $S^2$ is conical.       

The second result concerns imposing the equations that follow from the vacuum Einstein equations with cosmological constant in the case when the horizon is (locally) embedded in such spacetime and the Weyl tensor is of the Petrov Type D at the horizon.  
We have derived all the solutions, and they are characterized by conically singular data defined on $S^2$ and constructed the corresponding horizons of the Hopf bundle structure and the null direction transversal to the fibers.
An a priori nonobvious conclusion is that every solution gives rise to a nonsingular horizon, provided the two-dimensional metric tensor has at most conical singularities at the poles - the corresponding rotation scalar emerges as a regular function.  

The third result concerns the Killing horizons contained in the accelerated vacuum Kerr-NUT-(anti-)de Sitter spacetimes with cosmological constant (both the acceleration and cosmological constant are allowed to vanish).  
We identified them with a subset of the family of the solutions we found. 
The subset is parametrized by the same number of free parameters as the set of the solutions, but we did not show 1-1 correspondence.  
While in the generic case of a Killing horizon in accelerated Kerr-NUT-(anti-)de Sitter spacetime the null direction is transversal to the fibers of the corresponding Hopf bundle, there are special cases when the fibers coincide with the null generators. 
They occur if the cosmological constant takes special values determined by the other parameters characterizing the spacetime and horizon.   

As an additional, but not aside, result of our work, we have constructed spacetimes of the topology $H\times \mathbb{R}$, where $H\rightarrow S^2$ is the Hopf bundle, locally isometric to the accelerated vacuum Kerr-NUT-(anti-)de Sitter spacetimes with cosmological constant, for every value of the parameters.  
We found that when the acceleration parameter is not zero, then the conical singularity can also be removed whenever the NUT parameter also does not vanish.

\noindent{Acknowledgments:}  
J. L. and M. O.’s research leading to these results has
received funding from the Norwegian Financial
Mechanism 2014-2021 UMO-2020/37/K/ST1/02788. M.
O. and D. D-R. were also supported by the Polish NCN
grant for the project OPUS 2021/43/B/ST2/02950.

\section*{Appendix A}
The special solution to the Petrov Type D equation  obtained in Sec \ref{sec:special-solution} may be parametrized solely by area radius $R$. 
Therefore, to investigate its embeddability we consider the Taub-NUT-(anti-)de Sitter metric in Eddington-Finkelstein coordinates:
 \begin{align}\label{TNUT}
     ds^2&= -\frac{Q}{r^2+l^2}\big(dv+2l(\cos\theta-1) d \phi \big)^2+2\big(dv+2l(\cos\theta-1)d \phi\big)dr+(r^2+l^2) \Big(d\theta^2+\sin^2\theta d\phi^2 \Big), 
 \end{align}
 where:
 \begin{align}
     Q = r^2 - 2Mr-l^2-\Lambda (-l^4+2l^2r^2+\tfrac{1}{3}r^4).
 \end{align}
 Since in contrast to our solution, the above metric is not well-defined at $\theta=\pi$, we are first obliged to perform Misner's gluing. Therefore, we apply the following transformation:
 \begin{align}
   v' = v-4l\phi, && r'=r, && \phi'=\phi, && \theta'=\theta, && \text{for} \  \ \ \theta,\theta'\neq 0, \pi.
 \end{align}
 The resulting metric is of the form
 \begin{align}
     ds'^2=& -\frac{Q}{r'^2+l^2}\big(dv'+2l(\cos\theta'+1) d \phi' \big)^2+2\big(dv'+2l(\cos\theta'+1)d \phi'\big)dr'\\
     \nonumber &+(r'^2+l^2) \Big(d\theta'^2+\sin^2\theta' d\phi'^2 \Big). 
 \end{align}
 Notice, that now the term $(\cos(\theta')+1)d\varphi'$ is regular at $\theta'=\pi$.
 Furthermore, the coordinate $v$ (and $v'$) becomes periodic with a period of $8\pi{l}/{m}$, where $m$ is an arbitrary integer. 
A priori $m$ is just an integer, but, in fact turns, out to be the topological charge, characterizing all $U(1)$ bundles. 
 Next, we consider the Killing horizons, obtained by the roots $r_H$ of the expression $Q=0$, that is
 \begin{align}
     r_H^2-2Mr_H-l^2-\Lambda(-l^4+2l^2r_H^2+\tfrac{1}{3}r_H^4)=0.
 \end{align}
Each of such horizons, as long as it is nonextremal, is one of the Type D horizon that we consider. 
 The 2-metric on a spacelike slice $S$ of the Killing horizon defined by $r=r_H$ is of the form:
 \begin{align}
     g_{AB}dx^Adx^B&= (r_H^2+l^2) \Big(d\theta^2+\sin^2\theta d\phi^2 \Big).
 \end{align}
 The coordinates $x$ and $\varphi$ are related to $\theta$, $\phi$ via
 \begin{align}
    x(\theta) = -\cos\theta, && \varphi (\phi) = \phi,
 \end{align}
 whereas the area radius may be expressed in terms of Taub-NUT-(anti-)de Sitter parameters:
 \begin{align}
     R^2=r_H^2+l^2.
 \end{align}
On the horizon, the Killing vector field
\begin{align}
    \xi = {4l}\frac{\partial}{\partial v}
\end{align}
defines the generator of the null symmetry, that is
\begin{align}
    \ell = \xi_{|H}.
\end{align}

\bibliographystyle{apsrev4-2}
\bibliography{bibliography.bib}
\end{document}